\documentstyle[11pt,newpasp,twoside,psfig]{article}
\begin{document}

\title{Predicting the clustering of X-ray clusters}

\author{L. Moscardini, S. Matarrese, F. Lucchin}
\affil{Universit\`a di Padova,  Italy}

\author{S. De Grandi}
\affil{Osservatorio Astronomico di Brera, Merate (LC), Italy}

\author{P. Rosati}
\affil{European Southern Observatory, Garching, Germany}

\begin{abstract}
We present a theoretical model which aims at predicting the clustering
properties of X-ray clusters in flux-limited surveys for different
cosmological scenarios. The model uses the theoretical and empirical
relations between mass, temperature and X-ray luminosity, and
fully accounts for the redshift evolution of the underlying dark
matter clustering and cluster bias factor. We apply the model to the
RASS1 Bright Sample and to the XBACs catalogue.  The results show that
the Einstein-de Sitter models display too low a correlation length,
while models with a matter density parameter $\Omega_{\rm 0m}=0.3$
(with or without a cosmological constant) are successful in
reproducing the observed clustering.
\end{abstract}

\section{The model}

Matarrese et al. (1997) developed an algorithm to describe the clustering in
our past light-cone, where the non-linear dynamics of the dark matter
distribution and the redshift evolution of the bias factor are taken
into account (see also Moscardini et al. 1998, 1999a,b and
Suto et al. 1999). The observed spatial correlation function $\xi_{\rm
obs}$ in a given redshift interval ${\cal Z}$ is given by the exact
expression
\begin{equation}
\xi_{\rm obs}(r) = { \int_{\cal Z} d z_1 d z_2 {\cal N}(z_1) r(z_1)^{-1} 
{\cal N}(z_2) r(z_2)^{-1} ~\xi_{\rm obj}(r;z_1,z_2) \over 
\bigl[ \int_{\cal Z} d z_1 {\cal N}(z_1) r(z_1)^{-1} \bigr]^2 } \;,
\label{eq:xifund}
\end{equation} 
where $\xi_{\rm obj}(r,z_1,z_2)$ is the correlation function of pairs
of objects at redshifts $z_1$ and $z_2$ with comoving separation
$r$ and ${\cal N}(z)$ is the actual redshift distribution of
the catalogue.
 
An accurate approximation for $\xi_{\rm obj}$ over the scales
considered here is
\begin{equation} 
\xi_{\rm obj}(r,z_1,z_2) \approx b_{\rm eff}(z_1) b_{\rm eff}(z_2) 
\xi_{\rm m}(r,z_{\rm ave}) \;,
\label{eq:xifund}
\end{equation}
where $\xi_{\rm m}$ is the dark matter covariance function and $z_{\rm ave}$
is an intermediate redshift between $z_1$ and $z_2$.

The effective bias $b_{\rm eff}$ appearing in the previous equation
can be expressed as a weighted average of the `monochromatic' bias
factor $b(M,z)$ of objects of some given intrinsic property $M$ (like
mass, luminosity, ...), as follows
\begin{equation} 
b_{\rm eff}(z) \equiv {\cal N}(z)^{-1} \int_{\cal M} d\ln M' ~b(M',z) 
~{\cal N}(z,M')\, ,
\label{eq:b_eff}
\end{equation}
where ${\cal N}(z,M)$ is the number of objects actually present in the
catalogue with redshift in the range $z,~z+dz$ and $M$ in the range
$M,~M+dM$, whose integral over $\ln M$ is ${\cal N}(z)$. In our
analysis of the two-point correlation function for X-ray selected
clusters we will use for ${\cal N}(z)$ in eq.(\ref{eq:xifund}) the
observed one, while in the theoretical calculation of the effective
bias we will take the ${\cal N}(z,M)$ predicted by the model described
below. This phenomenological approach is self-consistent, in that our
theoretical model for ${\cal N}(z,M)$ will be required to reproduce
the observed cluster abundance and their $\log N$--$\log S$ relation.

For the cluster population it is extremely reasonable to assume that
structures on a given mass scale are formed by the hierarchical
merging of smaller mass units; for this reason we can consider
clusters as being fully characterized at each redshift by the mass $M$
of their hosting dark matter haloes. In this way their comoving mass
function can be computed using an approach derived from
the Press-Schechter technique. Moreover, it is possible to adopt for
the monochromatic bias $b(M,z)$ the expression which holds for
virialized dark matter haloes (e.g. Mo \& White 1996).  Recently, a
number of authors have shown that the Press-Schechter relation does
not provide an accurate description of the halo abundance both in the
large and small-mass tails.  Also, the simple Mo \& White (1996) bias
formula has been shown not to correctly reproduce the correlation of
low mass haloes in numerical simulations. We adopt the relations
recently introduced by Sheth \& Tormen (1999), which have been shown
to produce an accurate fit of the distribution of the halo populations
in the GIF simulations.

The last ingredient entering in our computation of the correlation
function is the redshift evolution of the dark matter covariance
function $\xi_{\rm m}$. We use an accurate analytical method to evolve
$\xi_{\rm m}$ into the fully non-linear regime. In particular, we use
the fitting formula given by Peacock \& Dodds (1996).

In order to predict the abundance and clustering of X-ray selected
clusters in flux limited surveys we need to relate X-ray cluster
fluxes into a corresponding halo mass at each redshift. The given band
flux $S$ corresponds to an X-ray luminosity $L_X=4\pi d_L^2 S$ in the
same band, where $d_L$ is the luminosity distance. To convert $L_X$
into the total luminosity $L_{\rm bol}$ we perform band and bolometric
corrections by means of a Raymond-Smith code, where an overall ICM
metallicity of $0.3$ times solar is assumed. We translate the cluster
bolometric luminosity into a temperature, adopting the empirical
relation $T = {\cal A} \ L_{\rm bol}^{\cal B} \ (1+z)^{-\eta}$, where
the temperature is expressed in keV and $L_{\rm bol}$ is in units of
$10^{44} h^{-2}$ erg s$^{-1}$. In the following analysis we assume
${\cal A}=4.2$ and ${\cal B}=1/3$; these values allow a good
representation of the local data for temperatures larger than $\approx
1$ keV.  Moreover, even if observational data
are consistent with no evolution in the $L_{\rm bol}-T$ relation out
to $z \approx 0.4$, a redshift evolution described by the parameter
$\eta$ has been introduced to reproduce the observed $\log N$--$\log
S$ relation (Rosati et al. 1998; De Grandi et al. 1999) in the range
$2 \times 10^{-14} \le S \le 2 \times 10^{-11}$.

Finally, with the standard assumption of virial isothermal gas
distribution and spherical collapse, it is possible to convert the
cluster temperature into the mass of the hosting dark matter halo (see
e.g. Eke et al. 1996).

\section {Results for RASS1 Bright Sample and XBACs}

We applied our method to different flux-limited surveys. Here we show
only the results obtained for the RASS1 Bright Sample (De Grandi et
al. 1999) and for the XBACs catalogue (Ebeling et al. 1996). The
application to other surveys (BCS, REFLEX and possible future space
missions) is presented in Moscardini et al. (1999b).

\begin{figure}
\centering  
\psfig{figure=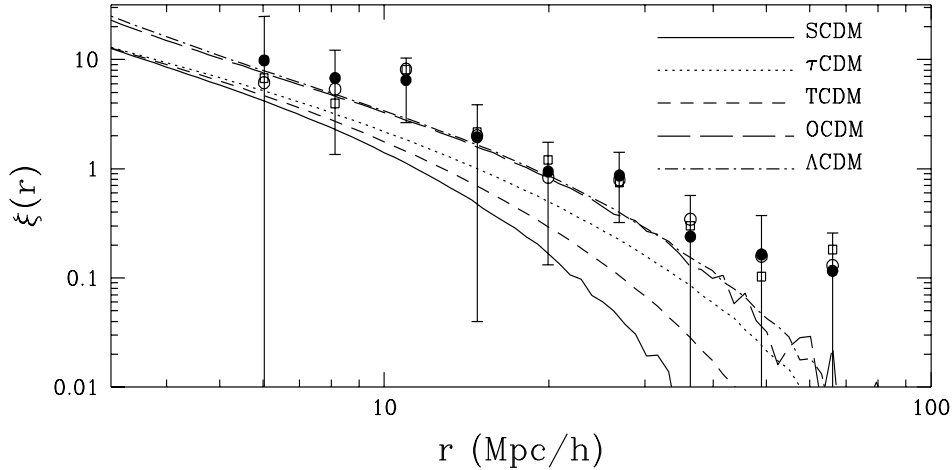,height=7.4cm,width=13cm,angle=0}
\caption{Comparison of the observed spatial correlation for clusters in the
RASS1 Bright Sample (Moscardini et al. 1999a) with the predictions of
the various theoretical models. }
\label{fi:theor}
\end{figure}

The RASS1 Bright Sample contains 130 clusters of galaxies selected
from the first processing of the $ROSAT$ All-Sky Survey (RASS1) data.
This sample was constructed as part of an ESO Key Programme aimed at
surveying all southern RASS candidates, which is now known as the
REFLEX cluster survey.  The RASS1 Bright Sample is count-rate-limited
in the $ROSAT$ hard band (0.5 -- 2.0 keV), so that due to the
distribution of Galactic absorption its effective flux limit varies
between 3.05 and $4\times 10^{-12}$ erg cm$^{-2}$ s$^{-1}$ over the
selected area. This covers a region of approximately 2.5 sr within the
Southern Galactic Cap. In Figure~\ref{fi:theor} we compare our
predictions for the RASS1 spatial correlation function in different
cosmological models to the observational estimes. Moscardini et
al. (1999a) find that the two-point correlation function $\xi(r)$ of
the RASS1 Bright Sample is well fitted by the power-law
$\xi=(r/r_0)^{-\gamma}$, with $r_0= 21.5^{+3.4}_{-4.4}\ h^{-1}$ Mpc
and $\gamma=2.11^{+0.53}_{-0.56}$ (95.4 per cent confidence level with
one fitting parameter). We considered five models, all normalized to
reproduce the local cluster abundance (Eke et al. 1996) and belonging
to the general class of Cold Dark Matter (CDM) models (the first three
are Einstein-de Sitter models): a standard CDM (SCDM) model; the
so-called $\tau$CDM model; a tilted model (TCDM), with spectral index
$n=0.8$ and with a high (10 per cent) baryonic content; an open CDM
model (OCDM), with matter density parameter $\Omega_{\rm 0m}=0.3$ and
a low-density flat CDM model ($\Lambda$CDM), with $\Omega_{\rm
0m}=0.3$.  All the Einstein-de Sitter models here considered predict
too small an amplitude. Their correlation lengths are smaller than the
observational results: we find $r_0\simeq 11.5, 12.8, 14.8 \ h^{-1}$
Mpc for SCDM, TCDM and $\tau CDM$, respectively.  On the contrary,
both the OCDM and $\Lambda$CDM models are in much better agreement
with the data and their predictions are always inside the 1-$\sigma$
errorbars ($r_0\simeq 18.4, 18.6 \ h^{-1}$ Mpc, respectively).

\begin{figure}
\centering  
\psfig{figure=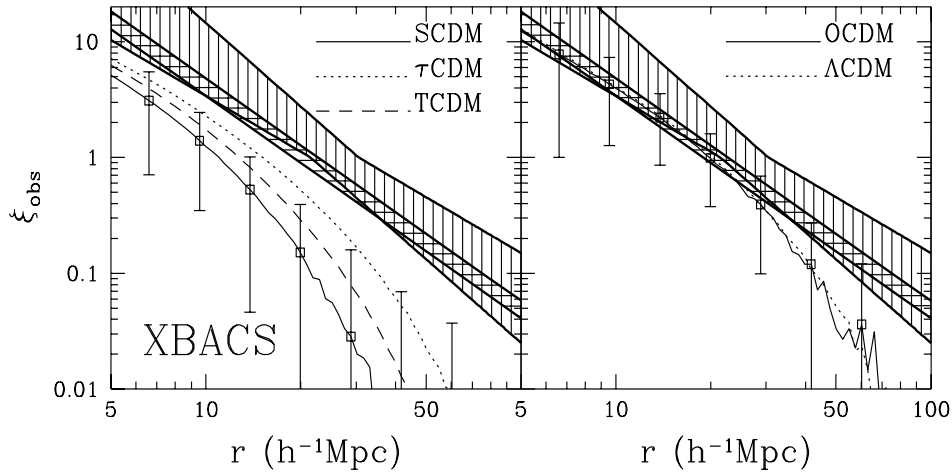,height=7.4cm,width=13cm,angle=0}
\caption{Comparison of the observed spatial correlation for clusters in
the XBACs sample with the predictions of the various theoretical
models.  The horizontal shaded area refers to the (1-$\sigma$)
observational estimates obtained by Abadi et al.(1998), the vertical
shaded one shows the (2-$\sigma$) estimates by Borgani et
al. (1999). }
\label{fi:xbacs}
\end{figure}

The XBACs catalogue is an all-sky X-ray sample of 242 Abell galaxy
clusters extracted from the ROSAT All-Sky Survey data. Being optically
selected, it is not a complete flux-limited catalogue.  The sample
covers high Galactic latitudes ($|b_{II}|\ge 20^o$). The adopted
limiting flux is $S_{\rm lim}=5 \times 10^{-12}$ erg cm$^{-2}$
s$^{-1}$ in the 0.1--2.4 keV band.  Due to the aforementioned
selection effects, the XBACs luminosity function $N(L)$ in the faint
part is much lower than that obtained from other catalogues. Using a
redshift evolution of the temperature-luminosity relation, we forced
our models to be consistent with the number counts. For this reason we
have to introduce in the models for XBACs its incompleteness $I(L)$,
defined as the ratio between its luminosity function $N_{\rm
XBACs}(L)$ and $N_{\rm real}(L)$, which is a combination of the
results for RDCS at low $L$ (Rosati et al. 1998) 
and for BCS at high $L$ (Ebeling et al. 1997).  The clustering
properties of this catalogue have been studied by different authors.
Abadi et al. (1998) found that $\xi(r)$ can be fitted by the usual
power-law relation with $\gamma=1.92$ and $r_0=21.1^{+1.6}_{-2.3}\
h^{-1}$ Mpc (errorbars are 1 $\sigma$). Borgani et al. (1999), who
adopted an analytical approximation to the bootstrap errors, found
$\gamma=1.98^{+0.35}_{-0.53}$ and a slightly larger value of
$r_0=26.0^{+4.1}_{-4.7}\ h^{-1}$ Mpc (errorbars in this case are
2-$\sigma$ uncertainties). Figure~\ref{fi:xbacs} compares these
observational estimates to the theoretical predictions of the
cosmological models previously introduced.  Again we find that all
Einstein-de Sitter models display too a small clustering.  Their
correlation lengths are smaller than the observational results: we
find $r_0\simeq 11, 15, 13\ h^{-1}$ Mpc for SCDM, TCDM and $\tau CDM$,
respectively.  On the contrary, both the OCDM and $\Lambda$CDM models
give very similar results and are in better agreement with the
observational data ($r_0 \simeq 20-22 \ h^{-1}$ Mpc).

\section{Conclusions}
We believe that the method presented here leads to robust predictions
on the clustering of X-ray selected galaxy clusters. Its future
application to new and deeper catalogues will provide a useful
complementary tool to the traditional cluster abundance analyses to
constrain the cosmological parameters.

\end{document}